# Scattering-assisted and logic-controllable WGM laser in liquid crystal micropillar

Jin Chuan Zhang, Hong Yang Zhu, Xiao Mei Zhu, Yan Li Zhang, Zhao Wang, Fei Liang Chen, Ke Li, Xiaofeng Li, and Wei Li Zhang*,

*Abstract*—**Whispering gallery mode (WGM) microcavities can efficiently store and manipulate light with strong light confinement and long photon lifetime, while coupling light into and from WGMs is intrinsically hindered by their unique feature of rotational symmetry. Here, a scattering-assisted liquid crystal (LC) micropillar WGM laser is proposed. WGM lasing at the surface of the micropillar is obviously enhanced by fluorescence scattering in the core of the micropillar. Besides, weak scattering of LC molecules also builds efficient coupling channels between the laser modes and the axial transmission modes of the micropillar-based waveguide, providing an all-in-one liquid WGM laser with functions of self-seeding and self-guiding. Furthermore, based on the hysteresis characteristics of the electrically anchored LC molecules under the interaction of thermal force, an erasable read-write liquid memory device is proposed, paving the way for the application of logic-controllable WGM lasers in optical storage and optical control.**

## I. INTRODUCTION

Optical microcavities provide a good platform to confine light and mediate light-matter interaction [1-3], which have arisen extensive research interest in areas of laser generation, frequency comb and functional photonic devices, serving various applications such as optical integrating, imaging, sensing and information technology [4-10]. In microstructures, like micro-rings, micro-disks, microspheres and microplanars [11-14], Whispering gallery mode (WGM) or morphology-dependent resonances occur in walls of such microcavities where light can be guided around and circulate due to total internal reflection. WGMs arise in structures with a higher refractive index compared to the environment and when the circulating light meets the phase matching condition. They usually have small mode volumes and very high Q-factor, which can well fit into the integrated photonics as miniaturized optical components such as laser sources, active filters, and all-optical switches [15,16]. In addition, the use of microcavity lasers to construct pixelated RGB arrays can realize full-color laser display, which provides a feasible solution and new approach to develop high-performance, easy-to-process, flat-panel or three-dimensional display devices [17,18].

Finding better ways to operate WGMs, e.g., coupling and tuning the output, is a flourishing research direction. By changing different stimuli, such as size, mechanical shape, temperature or electric field, free spectral range (FSR) and Q-factor tuning have been reported [19-21]. Due to the good symmetry of the cavity boundary, the photons are well trapped inside the cavity, and the mode cannot be excited or collected from outside, which limits the integration and application of the device to a certain extent. Thus, the efficient coupling has become an important prerequisite to operating WGMs. Evanescent field coupling using nano/micro-waveguide has become a primary method to collect/excite WGMs, however, which requires strict phase matching condition and fine alignment. In recent years, various cavities, like spiral and triangular shape cavities [22,23], have been proposed to support directional emission and easier output coupling. Deformed cavities and new coupling mechanisms have also been investigated to relax restrictions on the exciting and coupling of WGMs [24-26].

Currently, hollow-core fibers (HOFs) have become a popular platform for generating WGMs, as they can be mass-produced and can be easily combined with optofluidic technology [27,28]. In WGM lasers made of HOFs filled with gain media, the gain media typically have lower refractive index than the walls of the HOFs. Thus, precise processing is necessary to enable HOFs with very thin and highly finessed wall structures to support the lasing of WGMs [29,30]. The output guidance of laser modes also needs strict matching in momentum space [12,14].

Here, a scattering-assisted liquid crystal (LC) micropillar WGM laser is proposed, wherein weak scattering of the LC molecules is used to couple light between the core and the surface of the micropillar, providing an efficient mechanism for enhancing the WGM lasing at the surface and coupling the lasing into the axial direction of the micropillar. The LC micropillar is formed by filling dye-doped LC solution into HOFs with different inner-core diameters. As the inner

Manuscript received *****, 2022. This work is supported by the National Natural Science Foundation of China under Grants 11974071, 61635005 and 61875132, and Natural Science Foundation of Jiangsu Province under Grant BK20180208.( Corresponding author: W.L. Zhang (wl_zhang@uestc.edu.cn)
J. C. Zhang and H. Y. Zhu contributed equally to this work.

J. C. Zhang, H. Y. Zhu, Z. X. Hu, X. M. Zhu, Y. L. Zhang, Z. Wang, W. L. Zhang are with Fiber Optics Research Centre, School of Information and Communication Engineering, University of Electronic Science & Technology of China, Chengdu 611731, China.
F. L. Chen is with School of Electronics Science Engineering, University of Electronic Science and Technology of China, Chengdu, 610054, China.
K. Li is with Wenzheng College of Soochow University, Suzhou 215104, China.
X. F. Li is with College of Physics, Optoelectronics and Energy & Collaborative Innovation Center of Suzhou Nano Science and Technology, Soochow University, Suzhou 215006, China.

medium has a higher refractive index than the walls of HOFs, WGMs will mainly be confined by the surface of the filler other than by the walls of the HOFs, requiring no restraint to the wall thickness of the HOFs, i.e., there is no need to constrain the wall thickness of HOFs by precise processing. Besides, the LC micropillar also performs as a fiber-type waveguide, which is assisted by weakly light scattering, for high-efficient output coupling of WGM laser. Thus, an all-in-one liquid WGM laser is proposed with functions of self-seeding and self-output-coupling. Moreover, a novel bistable WGM operation mechanism is proposed based on the hysteresis characteristics of the electrically anchored LC molecules under interaction of thermal force, paving a way for the application of logical-controllable liquid WGM lasers in optical memory and optical control.

## II. RESULTS AND DISCUSSION

### A. *Characteristics of the scattering-assisted WGM lasers*

The dye solution used in this experiment includes 99.5 % E7 LC, and 0.5 % laser dye Pyrromethene 597 (PM597 from Exciton) by weight ratio. The E7 LC is a eutectic mixture of nematic LC (see details in Fig. S1), which has large birefringence (refractive index $n_e = 1.7462$ for extraordinary light and $n_o = 1.5216$ for ordinary light) and wide nematic temperature range (from $-10°C$ to $59°C$) [31]. The LC mixture was heated to isotropy on a heating plate with stirring to ensure uniform dispersion of fluorescent molecules in the LC. The mixed solution was sucked in and filled into the cores of HOFs through siphoning, forming active LC micropillars. The schematic setup to excite WGM laser from the LC micropillars is illustrated in Fig. 1(a). A 532 nm Q-switched laser with a pulse duration of 6 ns and repetition frequency of 1 Hz is used as the excitation pump. In order to tune the state of the LC molecules, the HOF is sandwiched between two indium-tin oxide (ITO) coated glasses and placed on a temperature controller (Instec mK2000, 0.01-200°C/min). A multi-mode fiber (MMF) with a core diameter of 105 μm is used to collect the output light along the axial direction.

Due to the difference between the refractive indices of LC micropillars and the supporting HOFs, WGM resonances are supported at the surface of the LC micropillars based on total internal reflection. Scattering of LC molecules builds coupling channels between the surface and other regions of the LC micropillar. On one hand, more fluorescence will be scattered into the WGM modes, which performs as seed light and is amplified by stimulated emission to achieve lasing with balanced gain and loss. On the other hand, the laser emission can be better coupled into the core region of the micropillar with nonzero axial momentum, transforming the lasing modes into axial transmission modes. Then the LC micropillar performs as an optical waveguide that delivers the output of the WGM laser in the axial direction, providing better coupling to optical fibers and other devices. Thus, our WGM laser is inherently scattering-assisted and self-guided with enhanced emission efficiency and axial output directionality. While in traditional WGM lasers, light is excited and coupled in the tangential direction of the WGM profile through the strict phase matching condition.

The optical resonance in the WGM microcavity was numerically simulated using the Finite difference time domain (FDTD) method. A cylindrical ring structure was set up with a core diameter of 10 μm and an outer diameter of 300 μm. To well fit the experimental results, the refractive index of the core (LC) was set to 1.63 without considering the errors of cavity length, and the refractive index of the outer wall (SiO₂) was set as 1.46. A wideband electric dipole source with wavelength centered at 585 nm was placed near the interface between the core and the outer wall, i.e., the surface of the LC micropillar. Fig. 1(b) shows the resonance spectrum of the structure with a Q-factor about 650, wherein resonant modes with FSR around 6.7 nm are observed. The insets give power distribution of different resonance modes, and show well confined standing waves around the surface of the micropillar, indicating the formation of WGM modes.

In order to further verify the emission mechanism of the WGM laser, resonance characteristics of the LC micropillars with diameters from 10 μm to 150 μm are studied experimentally. The upper inset 1 of Fig. 1(c) shows the luminous states of the micropillars pumped by a 530 nm continuous-wave (CW) laser, which indicates that straight LC micropillars with different diameters are formed. The curve corresponds to the theoretical results for FSRs as a function of the core diameters, i.e., $\Delta\lambda = \lambda^2 / n_{eff}\pi D$ , where $\lambda$ is the resonant wavelength, $D$ is the diameter of the micropillar, and $n_{eff}$ is the effective refractive index of the lasing modes, which can be calculated as $n_{eff} = \varphi n_{ave} + (1-\varphi) n_{glass}$ , where

$$\varphi = \frac{\int_{cavity} |E|^2 \, dr}{\int_{total} |E|^2 \, dr}$$ is the fraction of power inside the microcavity,

$n_{ave}$ is the average refractive index of the filler, and $n_{glass}$ is the refractive index of the glass wall [32,33]. In our case, $\varphi$ is close to 1 because the WGMs are mainly confined within the core. Therefore, $n_{ave}$ is used to represent $n_{eff}$ for simplification, i.e., $n_{eff} \approx n_{ave} = (2n_o + n_e)/3$ in the theoretical analysis [34]. Due to the WGM modes circulate around a circular surface, and the light will go through LC molecules arranged in random directions, the effective refractive index of LC can always be the same average value, and there is no TM/TE mode related to the direction of the electric field oscillation [19].

The dots in Fig. 1(c) correspond to experimental results for LC micropillars of different sizes when pumped by the 532 nm pulsed laser, which agree with the theoretical results, demonstrating the lasing of WGMs supported by the LC micropillars. As shown by the spectra in the lower inserts of Fig. 1(c), only two separate peaks can be excited for $D$ =10 μm due to the large mode separation related to the small cavity length. The corresponding FSR is about 6.7 nm which is consistent with the simulated value in Fig. 1(b), and the Q-factor of the lasing modes is about 500. It is worth noting that more WGM modes are excited as $D$ increases (>50 μm), the laser linewidth is narrowed and the Q-factor increases to 2000. The fluorescence intensity generated in the core region of the micropillar also increases with the gain volume, which leads to the uplift of the fluorescent substrate of the collected output of the WGM lasers.

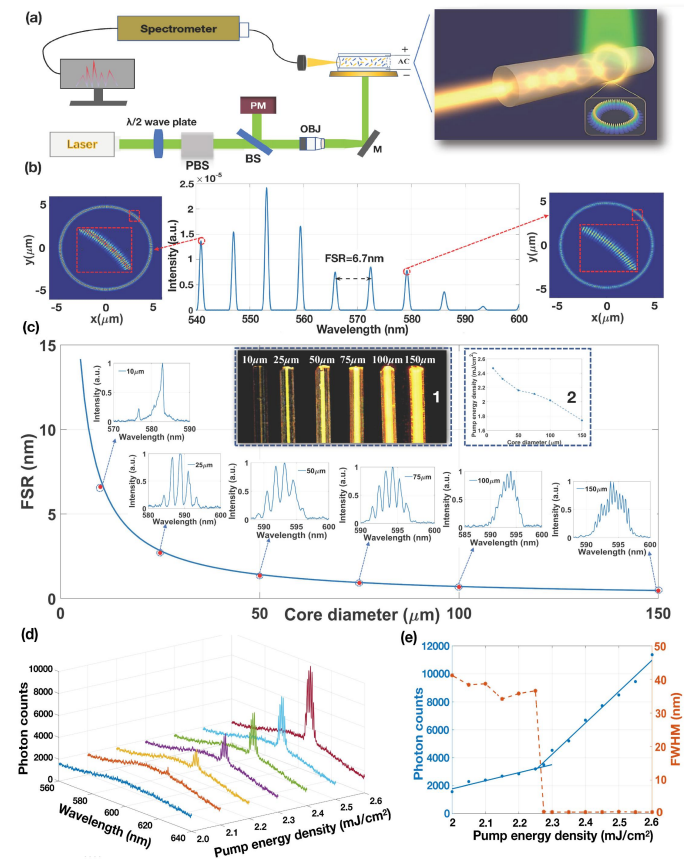

Fig. 1. Setup and output characteristics of the proposed WGM lasers. (a) Schematic setup of the WGM laser. PBS, polarization beam splitter. BS, beam splitter. PM, power meter. OBJ, microscope objective. The inset is a schematic diagram of the LC micropillar that generates the scattering-enhanced all-in-on WGM lasers. (b) Simulated resonance spectrum and power distribution of different resonant modes when $D = 10$ μm. (c) FSR as a function of $D$. The blue curve and the red dot represent theoretical and experimental values, respectively. The upper inset 1 is the fluorescence image of the micropillars and the lower insets are the output spectra of the WGM lasers. The upper inset 2 is the threshold of WGM lasing as a function of $D$ (d) Output spectra at different values of pump intensity when $D = 75$ μm. (e) Output power (blue) and linewidth (red) of the lasing peak as functions of pump intensity when $D = 75$ μm.

Fig. 1(d) and 1(e) show the lasing of WGMs when $D = 75$ μm. The output spectra show multiple peaks with equal mode separations (e.g., FSR of ~0.94 nm) and enhanced peak intensity when increasing the pump intensity. The threshold of lasing is about 2.2 mJ/cm². When pump intensity is below the threshold (<2.2 mJ/cm²), the output spectrum is a fluorescent substrate with a bandwidth of about 40nm and is mainly determined by spontaneous emission. A kink (i.e. laser threshold) is presented in the input–output curve at the pump energy density of 2.2 mJ/cm², indicating the spontaneous emission converts to the stimulated emission. The linewidth of an individual laser mode is about 0.4 nm. With the increase of pump intensity (>2.2 mJ/cm²), the formation of a stable multi-peak laser can be clearly observed clearly, and the intensity and position of peaks do not change pulse to pulse, which are typical characteristics of WGM modes. The WGM lasing threshold decreases with increasing core diameter, as shown in the upper inset 2 of Fig. 1(c). The WGM lasing threshold decreases with increasing of the core diameter, as shown in the upper inset 2 of Fig. 1(c). In the WGM laser regime, the micropillars with larger core diameters have larger gain regions, which may reduce the lasing threshold. Besides, the scattering effect plays a positive role in WGM lasing. In a larger micropillar, more fluorescence scattering might couple into the WGMs to provide more seed light that helps to reduce the lasing threshold.

The shape of output spectrum keeps almost unchanged when changing the pump position along the axial direction of the micropillar, as shown in Fig. 2(a), which is attributed to the high structural uniformity of the LC micropillar and is helpful to build arrays of identical WGM lasers. Besides, Dozens of samples were made in the experiment, and the samples of the same size can have very similar WGM emission. As shown in Fig. 2(b), the typical spectra of five different samples are given when $D = 75$ μm, indicating the repeatability of the experimental results.

The waveguide effect of the LC micropillar was also studied using the FDTD method. In Fig. 2(c), a core diameter of 10 μm is considered and a Gaussian light source centered at 590 nm was placed at center of the core with axial propagation direction. It shows that the light is well confined in the core and propagates along the axis of LC micropillar. In contrast, when the ethanol solution with lower refractive index (about 1.36) was filled into the core by simulation, the waveguide effect of the laser beam was not significant, and most of the light leak out of the wall, as shown in Fig. 2(d).

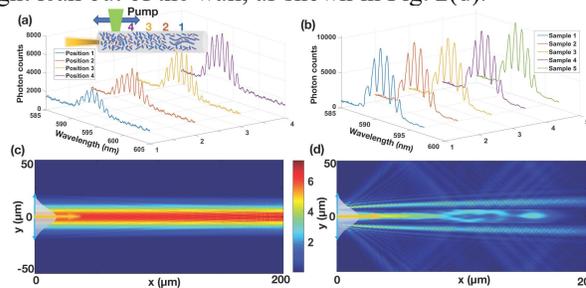

Fig. 2. Output and coupling characteristics of the proposed WGM lasers. (a) The output spectrum along the axial direction of the micropillar when changing the pump position and $D = 10$ μm (b) Output spectra of different samples when $D = 10$ μm. (c) and (d) The Simulated waveguide effect of micropillar when filled with LC eutectic mixture and ethanol solution.

To better understand the role of scattering in WGM lasing, the LC micropillar was heated to different values of temperature with accuracy of 0.1 °C considering the dynamic scattering mechanism of LC molecules. The core diameter size of the LC micropillar was chosen as 75μm because of moderated value of FSR, medium number of WGMs, and high contrast between lasing modes and the background fluorescence. It is known that temperature determines the internal energy of the flow of LC molecules, which causes pronounced reorientation of molecules and the optic axis. Increasing temperature causes expansion of molecules and spontaneous fluctuations of the nematic directors, which decreases the order of the molecules and induces a variation in the dielectric tensor due to the strong gradients of concentration of the cyanobiphenyl components in the LC mixture. Eventually, scattering effect related to the arrangements of LC molecules increases significantly [35,36]. Correspondingly, the lasing threshold of the LC micropillar decreases with increasing temperature, as shown in Fig. 3(a), which reflects the positive role of light scattering in the WGM laser emission, i.e., more fluorescence scattering is coupled into the WGM modes to provide more seed light that helps to reduce the lasing threshold. From Fig. 3(a), the slope efficiency of the WGM laser slightly decreases with temperature increasing, mainly because that thermal-dynamic accumulation reduces the fluorescence efficiency of the gain medium [37]. This is also verified in Fig. S2, where the dye is dissolved in a non-scattering solution to eliminate the influence of the scattering effect, and the fluorescence intensity decreases with temperature increasing. However, the enhancement of the thermal-induced scattering-assisted lasing outweighs the decrease of the thermal-induced fluorescence and plays a dominant role. As a result, the output intensity of the WGM laser increases with temperature at fixed pump intensity. Besides, below the phase transition temperature, the average refractive index of LC decreases slightly with the increase of temperature (as shown in Fig. S3), which is not conducive to the formation of WGM. This also reflects that the increase in scattering causes the enhancement of WGM lasing.

The mechanism of scattering-assisted WGM laser can be phenomenological analyzed through a rate equation model [38]:

$$\frac{dP(t)}{dt} = gP_g P(t) - \gamma P(t) + \eta P_{ASE}(t) \tag{1}$$

The first term of the right part of Eq. (1) represents the contribution of the pump and the gain, the second term represents the total loss of the laser cavity, and the third term represents the contribution of the seed light from spontaneous emission. Symbol $P$ represents the output of the WGM laser, and $P_g = (P_{in} - P_{th})/P_{th}$ is the normalized intensity of the pump, wherein $P_{in}$ is the pump intensity and $P_{th}$ is the pump intensity at the lasing threshold. Symbols $g$ is the gain index, $\gamma$ is the decay rate, $\eta$ is the coupling rate from the whole fluorescence to the laser modes and it is positively proportional to the scattering strength. For the average intensity of the laser, Eq. (1) can be simplified to a steady-state equation independent of time:

$$gP_g P - \gamma P + \eta P_{ASE} = 0 \tag{2}$$

Taking the curve at 30 °C as an example, the relationship between $g$ and $\eta$ can be found clearly. When $P_{in} = P_{in0}$ (as marked in Fig. 3(a), $P_{in0}$ is the pump intensity at the inflection point), we have $P = P_{ASE}$ and can get $\gamma = g(P_{in0} - P_{th})/P_{th} + \eta$. As $\gamma$ depends little on the pump intensity and temperature, it is assumed to be constant. Replacing $\gamma$ by $g(P_{in0} - P_{th})/P_{th} + \eta$ in Eq. (2), the following relationship can be determined:

$$\frac{\eta}{g} = \frac{P(P_{in} - P_{in0})}{P_{th}(P - P_{ASE})} \tag{3}$$

which reflects the coupling coefficient from the scattered fluorescence relative to the gain. From the experimental results, values of the parameters in the right part of Eq. (3) can be determined, as marked in Fig. 3(a), and $\eta/g$ can be calculated. Thus, the relative coupling coefficient $\eta/g$ versus temperature relationship can be found by repeating the above procedure at different temperatures. Fig. 3(b) gives the values of $P$ and $P_{ASE}$ as a function of temperature when $P_{in} = 2.4$ mJ/cm². The fluorescent output intensity, $P_{ASE}$, decreases with temperature due to thermal-induced fluorescence reduction. Oppositely, the laser output, $P$, increases with temperature due to thermal-induced scattering enhancement, indicating the positive role of scattering during WGM laser operation. Under WGM lasing regime, even though the fluorescence/gain efficiency $g$ decreases with temperature, Fig. 3(c) shows that $\eta/g$ increases significantly with temperature, indicating that the effect of scattering-assistant lasing becomes more dominant. The increased scattering provides more seed light to boost WGM lasing, which overcomes the influence of the decreasing fluorescence efficiency. Thus, this result verifies that our WGM laser works in a new regime, namely, scattering-assistance with enhanced emission intensity, which is different from the regimes of traditional WGM lasers (where almost no scattering takes effect) or random lasers (where the laser modes are solely determined by multiple scattering and no resonance of WGMs takes effect) [39-42].

When LC molecules reach the isotropy state above the phase transition temperature $T_c$ (>60°C), the utterly disordered state of the LC molecules in the micropillar enhances the effect of random scattering greatly. In this case, strong scattering-related random feedback overcomes the total internal reflection of the WGM modes, and the WGM lasing regime converts to a random lasing regime, as shown in Fig. S4 and the former reported works [39]. In our work, the mixture of LC with a relatively high phase transition threshold is used, and our operation temperature is below this threshold (<60°C), which ensures the scattering is always moderate to assist WGM lasing other than arousing random lasing.

Parallelly, an electrical field was applied to control the scattering effect of the LC micropillar. A 1000 Hz alternating current voltage, $U_{AC}$, was applied to the ITO glasses clamping the HOF, which deflects the LC molecules uniformly under the electric field force, reorientating the directors of LC molecules, i.e., the optical axis. As shown in the inset of Fig. 3(d), when $U_{AC} = 0$, the LC micropillars show a cloudy and bright state under the microscope, and the total output intensity in the axial direction has a relatively large value. This reflects a strong scattering state of the LC molecules because they are only weakly anchored by the inner wall of the HOF and are almost randomly oriented. As $U_{AC}$ increases, the electric force added to LC molecules gradually overcome the Fredericks transition threshold, and the molecules are deflected along the direction of the electric field. During this procedure, the LC micropillars become more transparent under the microscope and the total output intensity becomes weaker as more LC molecules align with the applied electric field, i.e., less disordered. Correspondingly, the lasing threshold of the LC micropillar increases with the increase of $U_{AC}$, as shown in Fig. S5, which follows the conclusion of Fig. 3(c), and also verifies that the scattering can enhance the emission efficiency and axial guidance of the WGM lasers. In the process of collecting the laser intensity, the change intervals for temperature and voltage are about 0.5 minutes to achieve stable molecular state, and the number of integrated photons does not change significantly with stable single pulse to obtain a stable laser output.

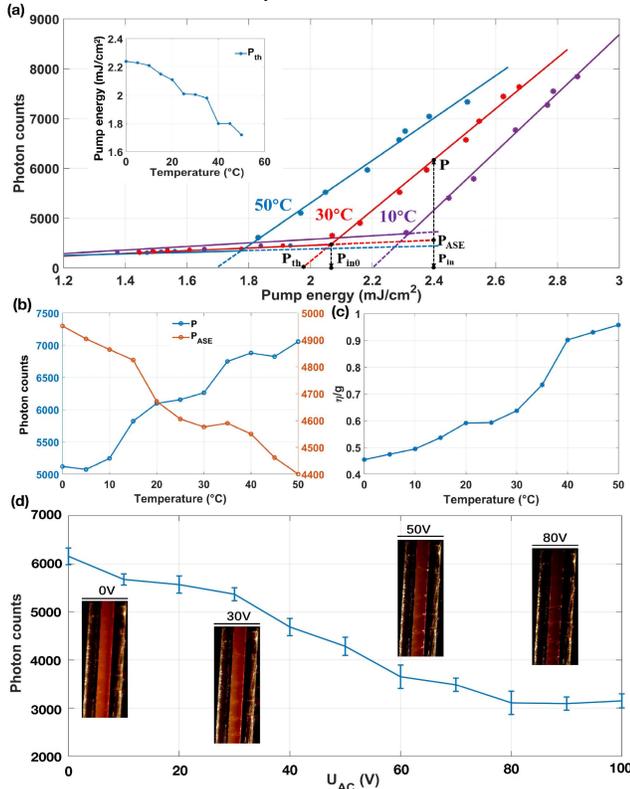

Fig. 3. Role of scattering in emission of the WGM laser, for $D = 75$ μm. (a) The input-output curves of the WGM laser at different values of temperature. The inset shows lasing threshold vs temperature. (b) Output intensity of the lasing and fluorescing regimes vs temperature, for $P_{in} = 2.4$ mJ/cm². (c) Variation of $\eta / g$ versus temperature. (d) Intensity variation of the WGM laser vs $U_{AC}$, for $P_{in} = 2.4$ mJ/cm² and temperature of 20 °C. The insets are the microscope images of the WGM laser at different $U_{AC}$.

## B. Logical operation of the WGM lasers

The above results show that the increasing voltage and temperature have the opposite effect on the order of the LC molecules. The interaction of LC molecules tends to be in balance under these external forces. Once the external force is changed, the overall arrangement of the molecules remains unchanged due to the influence of the intermolecular viscous force unless the external force is strong enough. This is the so-called memory/hysteresis effect of the LC molecules, i.e., the molecules have inertia to keep their previous state. Therefor, when scanning a control parameter in two opposite directions to change the molecules' state, the output would show bistability (co-existence of two states for the same control parameter value). Here, the hysteresis effects based on the combined effect of thermal and voltage [43,44] are exploited to build bistable WGM lasers with memorial and logical operation functions. Fig. 4(a) shows the variation in intensity of the WGM laser ($D = 75$μm and $P_{in} = 2.4$ mJ/cm²) versus temperature when different voltages, $U_{AC}$, are applied. For $U_{AC}$ = 0, the curves corresponding to the temperature undergoing a periodic sweep of $0° → 50° → 0°$ C have a high degree of overlap, i.e., there is no bistable effect. This is because the LC molecules are rarely anchored by the HOF. The order of LC molecules changes with temperature and is only affected by viscous resistance, which is not sufficient to maintain the previous state of LC molecules at the scanning temperature. For $U_{AC} = 25$V, the effect is almost the same. When $U_{AC}$ increases to 50 V, the optical axis of the LC molecules is deflected in the direction of the applied electric field. Such electrical force is strong enough to have an anchor effect over all the LC molecules and determines the final state of the LC molecules together with the thermal force. In this case, the LC molecules can well maintain the state related to the previous temperature, the so-called memorial effect of the previous state, showing a hysteresis/bistable characteristic when scanning the temperature in the opposite direction. The curves corresponding to the upward and downward scans of temperature have separated output intensity values from 5 to 30 °C, i.e., the hysteresis region, providing a bistable regime of the WGM laser which is of great importance for applications of optical memory and logical calculations.

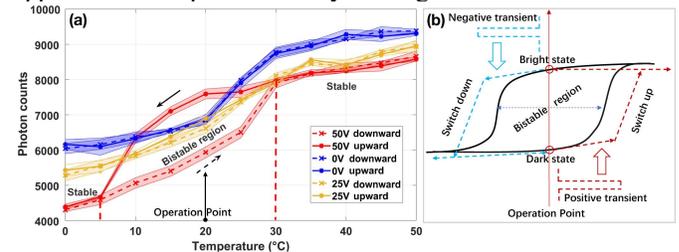

Fig. 4. Bistability of the WGM laser. (a) Intensity variation vs temperature at different values of $U_{AC}$. The red, yellow and blue color corresponds to $U_{AC} = 0$, 25V and 50V respectively. The symbols '*' and '×' corresponds to upward and downward scanning of temperature respectively. (b) Schematic diagram of logical operation.

The successive research demonstrate an application of logical operation based on our WGM lasers. Since the bistable effect appears when $U_{AC}$ is large enough, we use the condition of $U_{AC} = 50$ V as the switch and protection circuit for the logical device, e.g., the WGM lasers can perform as active read and write memories only when $U_{AC}$ is set to 50 V, which can avoid the misreading probability in data access. The state of the logical device can be prepared by a temperature that is equivalent to the "write" command in the digital storage device. The schematic diagram of the logical operation is shown in Fig. 4(b). For example, the operation temperature of the WGM lasers is set to 20 °C which is right within the bistable region, so that the laser can work at either the "dark" or the "bright" state. When a negative temperature signal is applied, i.e., a 20→0→20°C transient variation, the WGM laser will be set to the "dark" state. When a positive temperature signal is applied, i.e., a 20→50→20°C transient variation, the WGM laser will be set to the "bright" stat. Obviously, the "dark" and "bright" states can be coded as logical "0" and "1" respectively, and the LC micropillar can be written with logical information and read out by WGM lasing. It was also tested that the pump has negligible influence on the temperature of the sample, as shown in Fig. S6.

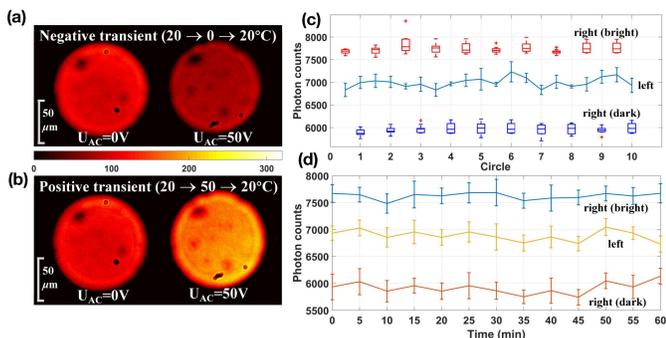

Fig 5. Logical operation of the WGM laser. (a)-(b) Output images of the two electrically and thermally controlled micropillars. The left/right one corresponds to $U_{AC}$ = 0/50 V, and (a)/(b) corresponds to the case that negative/positive temperature signal is applied. (c)-(d) Stability of the device versus repetition number and running time.

Such erasable and controllable memory functions can be flexibly expanded to arrays of active devices to support multi-bit operations. An example is given in Fig. 5(a) and 5(b), where two LC micropillars were separately electrically controlled and pumped simultaneously. The left one was set to the monostable/non-bistable regime, i.e., $U_{AC}$ = 0 at the operating temperature of 20°C. The right one was set to the bistable regime, i.e., $U_{AC}$ = 50 V at the operation temperature of 20°C. When the two WGM lasers were triggered by positive and negative temperature signals simultaneously, the output of the left one kept almost unchanged, while the output of the right one switched between the "dark" and "bright" states.

The reconfigurable characteristics and time-domain stability of the proposed logical WGM laser are investigated in Figs. 5(c) and 4(d), where the same operation of Figs. 5(a) and 5(b) were repeated and monitored over a relatively long period. In order to avoid photobleaching of dye caused by long-term pumping, the sample was pumped by only ten pulses at each detection time point. The laser can be made into an open channel (micro-fluid channel) through which the gain solution can be circulated and replaced easily to avoid photobleaching [45]. In order to make the LC molecules have a relatively stable relaxation state after the temperature change to obtain a stable laser output, that is, the number of integrated photons does not change significantly with stable single pulse, we set rate of the temperature controller to 2°C/s, and the cycle time for "dark" and "bright" states is 0.5min. The single-state of the left WGM laser as well as the "bright" and "dark" states of the right WGM laser show relatively high stability versus repetition numbers and running time, reflecting good repeatability and stability of our logical operations. Although the operation speed of the proposed logical WGM laser is not comparable with solid-state devices in photonic integrations, it breaks a way of actively controlling the logical states of liquid devices with the advantages of being easily deformable and replaceable. It provides efficient ways in applications of predefining the states of light sources (e.g., active nodes), laser coding and light control under fluid environment, which are useful for display, imaging and optical information applications.

## III. CONCLUSIONS

In summary, we realized a kind of LC micropillar-based WGM lasers that are integrated with the function of self-seeding and self-output-coupling. The proposed WGM laser emits from total internal reflection along the surface and is assisted by fluorescence scattering in the core of the micropillar, which has enhanced emission intensity and axially-guided output direction. The voltage and temperature significantly influence perturbation and deflection of the LC molecular directors, which ultimately affects the lasing state of the WGM laser. This endows our laser with enhanced output intensity, generating the so-called scattering-assisted WGM lasers different from traditional WGM lasers or random lasers. It also prompts us to develop the thermally controllable logical WGM laser under the combined effect of voltage anchoring. The results reported in this paper may offer a new route to the generation and output coupling of liquid WGM microcavity lasers, avoiding the complicated fabrication process of the cavity structure and precise phase matching. In addition, based on the bistable characteristics, an erasable read-write active logical device is proposed with memory effect, which may open a new way to develop functional liquid devices, like digital memories, optical read/write and active logical components in the fluid state.

**Jin Chuan Zhang**, is currently working towards a Master's degree in optical engineering at the Key Lab of Optical Fiber Sensing & Communications (Education Ministry of China), University of Electronic Science & Technology of China. He received a B.S. degree from North University of China in 2019. His current interests include multimode fiber.

**Hong Yang Zhu**, is currently working towards a Ph.D. degree in optical engineering at the Key Lab of Optical Fiber Sensing & Communications (Education Ministry of China), University of Electronic Science & Technology of China, Chengdu, China. He received a B.S. degree from North University of China in 2018. His research interests include novel fiber lasers and liquid crystal.

**Xiao Mei Zhu**, is currently working towards a Master's degree in optical engineering at the Key Lab of Optical Fiber Sensing & Communications (Education Ministry of China), University of Electronic Science & Technology of China, Chengdu, China. She received a B.S. degree from Southwest Jiaotong University of China in 2020. Her research interests include novel fiber lasers and liquid crystal.

**Yan Li Zhang,** received the B.S. degree from the Jimei University, Xiamen, China, in 2018. She is currently working towards the Ph.D. degree in optical engineering with the Key Lab of Optical Fiber Sensing and Communications, Education Ministry of China, Chengdu, China. Her current research interests include Intelligent fiber lasers and disordered systems.

**Zhao Wang,** received the B.Eng. degree from Jiang Su University, Jiangsu Zhenjiang, China, in 2018. He is currently working toward the Ph.D. degree in optical engineering with the Key Lab of Optical Fiber Sensing and Communications (Education Ministry of China), University of Electronic Science and Technology of China, Chengdu, China. His research interests include imaging through scattering medium, fiber based imaging, computational imaging.

**Fei Liang Chen,** received the B.Eng from Xi'an Jiaotong University and Ph.D. degree from Shanghai Institute of Technical Physics, Chinese Academy of Sciences. His research interests include semiconductor micro-nano optoelectronic devices, novel nano-air-channel high-frequency optoelectronic devices, optoelectronic fusion millimeter-wave technology, novel optical quantum encryption and authentication technology, etc.

**Ke Li,** associate professor of optoelectronic information science and engineering. She received Ph.D. degree from University of Science and Technology of China in 2013, and worked in Anhui Institute of Optics and Fine Mechanics, Chinese Academy of Sciences from 2012 to 2013. From 2013 to 2017, she worked as a postdoctoral fellow at the School of Optoelectronic Information Science and Engineering, Soochow University, and worked as a postdoctoral fellow at the Department of Physics, Imperial College London for one year.

**Xiao Feng Li,** (S'05–M'12–SM'12) received B.Eng. and Ph.D. (hons.) degrees in communication engineering from Southwest Jiaotong University, Chengdu, China, in 2002 and 2007, respectively. From 2007 to 2010, he was a Research Fellow with the School of Electrical and Electronic Engineering, Nanyang Technological University, Singapore. From 2010 to 2011, he was employed with Imperial College London, London, U.K., as a Research Associate. Since January 2012, he has been with Soochow University, Suzhou, China, as a Full Professor. He is currently the Dean of School of Optoelectronic Science and Engineering, Soochow University. He has authored ~ 200 journal papers with over 50 invited talks in conferences. His research interests include nano photovoltaics, surface-plasmon physics and devices, hot-electron photodetection, and optical sensing. He is a Senior Member of the Chinese Optical Society. He had been an Associate Editor for OSA Applied Optics and IEEE Photonics Journal from 2016 to 2021 and is currently the Editorial Member of several other journals.

**Wei Li Zhang,** (M'13–SM'13) received B.Eng. and Ph.D. degrees in communication engineering from Southwest Jiaotong University, China, in 2003 and 2008, respectively. From 09/2008 to 10/2010, he was working as a research fellow at the School of EEE, Nanyang Technological University, Singapore. He began to work at the University of Electronic Science and Technology of China as an associated professor since November 2010, and as a professor since 2016. His interests include semiconductor microcavity laser, optical fiber lasers, and optical integration. He has authored/coauthored over 100 papers published in refereed professional journals and national and international conferences and he filed 10 patents. Dr. Zhang is on the editorial board committee of Photonic Sensors.

# Supplementary Material for the manuscript:

## 1. Content of the LC mixture

The E7 LC used is eutectic mixture of nematic LC, composed of 57% 4-cyano-4'-n-pentyl-biphenyl (5CB), 21% 4-cyano-4'-nheptylbiphenyl (7CB), 16% 4-cyano-4'-n-octyloxy-biphenyl (8OCB), and 12% 4-cyano-4''-n-pentyl-p-terphenyl (5CT) by weight ratio [1,2].

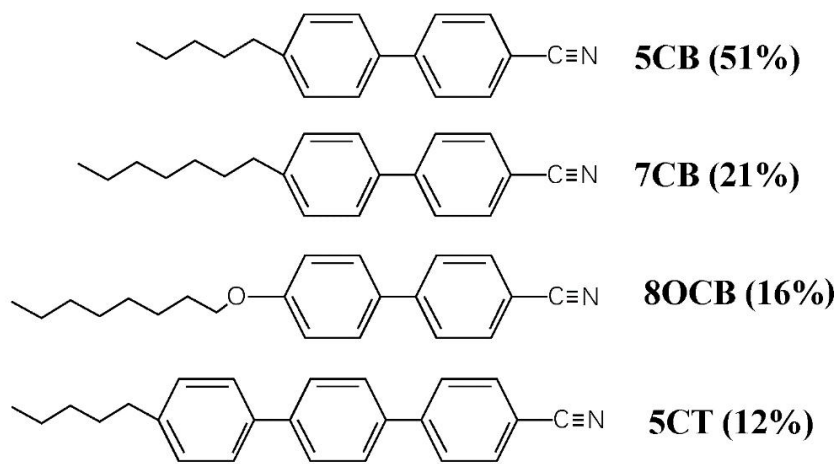

Fig. S1. The molecular formula and mass ratio of E7

## 2. Influence of temperature on fluorescence intensity of the dyes

The gain solution was prepared by dissolving 0.3 wt% Pyrromethene 597 dye powder in 99.8% ethanol, which was then filled into the core of the HOF to eliminate the influence of the scattering effect from the LC. The experimental results show that the fluorescence intensity in Fig. S2 decreases with temperature increasing.

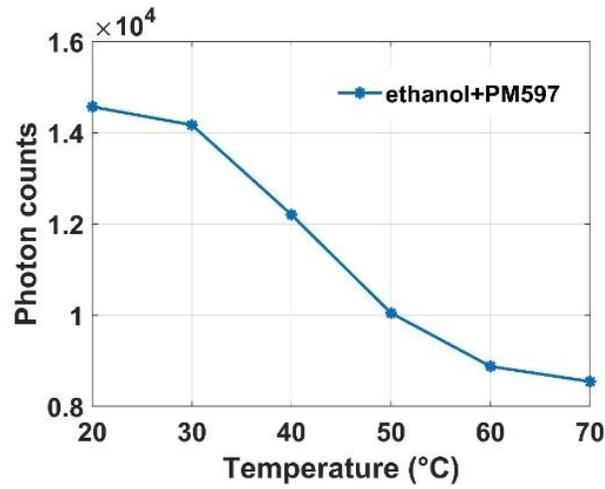

Fig. S2. Fluorescence intensity changes with temperature in PM597 ethanol solution. $P_{in}$ = 2.4 mJ/cm$^2$ and $D$ = 75 μm.

## 3. Influence of temperature on refractive index of the LC

The four-parameter model describing the temperature dependence of the LC refractive index is shown below [3,4]:

$$n_{ave}(T) = A - BT \qquad (1)$$

$$n_e(T) = A - BT + \frac{2(\Delta n)_0}{3}(1 - \frac{T}{T_c})^\beta \qquad (2)$$

$$n_o(T) = A - BT - \frac{(\Delta n)_0}{3}(1 - \frac{T}{T_c})^\beta \qquad (3)$$

In Eq. (1)-(3), A and B are the fitting constants for the specific wavelength. The average refractive index $n_{ave}(T)$ decreases linearly with increasing temperature, $(\Delta n)_0$ is the LC birefringence in the crystalline state (or $T$=0 K), the exponent $\beta$ is a material constant, and $T_c$ is the phase transition temperature of the LC. Fig. S3 depicts the temperature-dependent refractive indices of E7 at $\lambda$ =589 nm. Clearly, the average refractive index decreases slightly as the temperature increases.

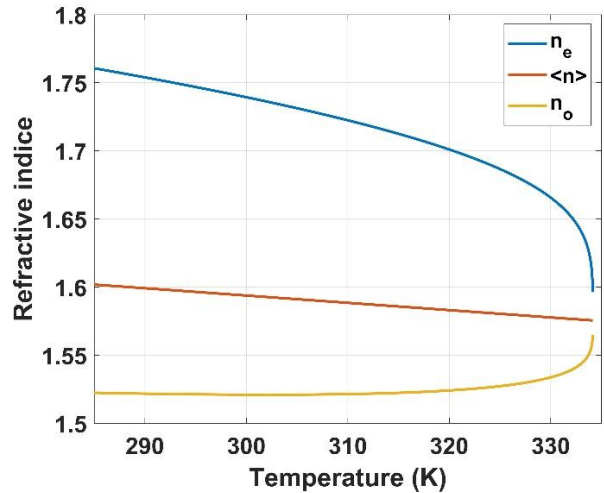

Fig. S3. Temperature-dependent refractive indices of E7 at $\lambda$ =589 nm.

## 4. Random lasing regime of the LC micropillar

When the LC molecules reach the isotropy state above the phase transition temperature (>60 °C), the completely disordered state of the LC molecules in the micropillar enhances the effect of random scattering greatly. In this case, strong scattering-related random feedback overcomes the total internal reflection of the WGM modes, and the WGM lasing regime converts to a random lasing regime. As shown in Fig. S4, the random laser spectra under 5 continuous single shots of the pump are completely different, indicating the LC micropillar works in the well-known coherent random lasing regime.

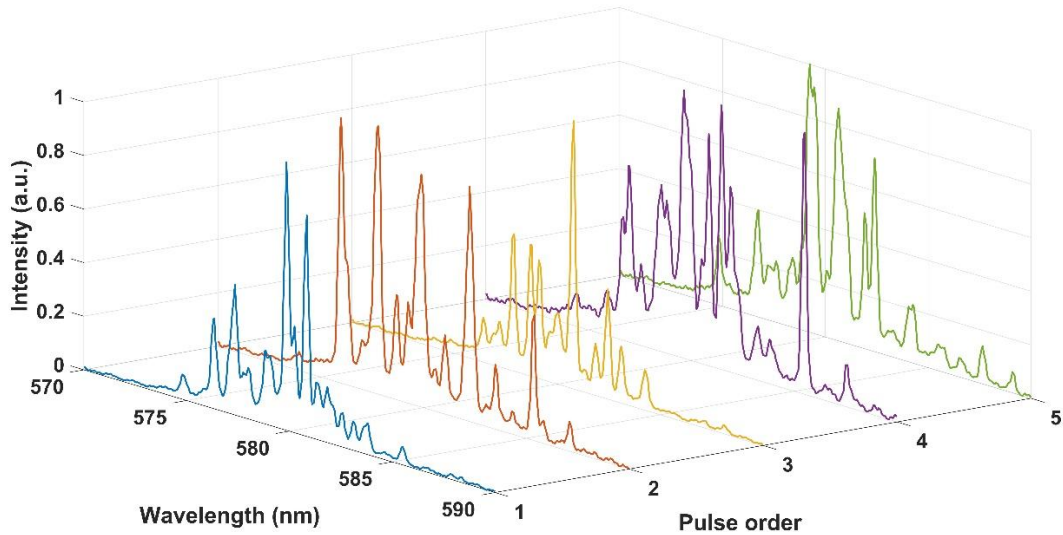

Fig. S4. Random laser spectra of liquid crystals in an isotropic state pulse to pulse. $P_{in} = 2.4$ mJ/cm$^2$ and $D = 75$ μm

## 5. Influence of voltage on the WGM laser

As $U_{AC}$ increases, the electrical force added to LC molecules gradually overcomes the Fredericks transition threshold, and the LC molecules deflect along the direction of the electric field. During this procedure, the LC molecules become more aligned to the applied electric field (i.e., less disordered). As a result, the threshold value of our scattering-assisted WGM laser becomes larger and larger (see Fig. S5), and its output intensity becomes weaker and weaker. This also reflects the positive role of scattering in the operation of our WGM lasers, and is consistent with the conclusion of the thermal-dependent scattering effect in Fig. 3.

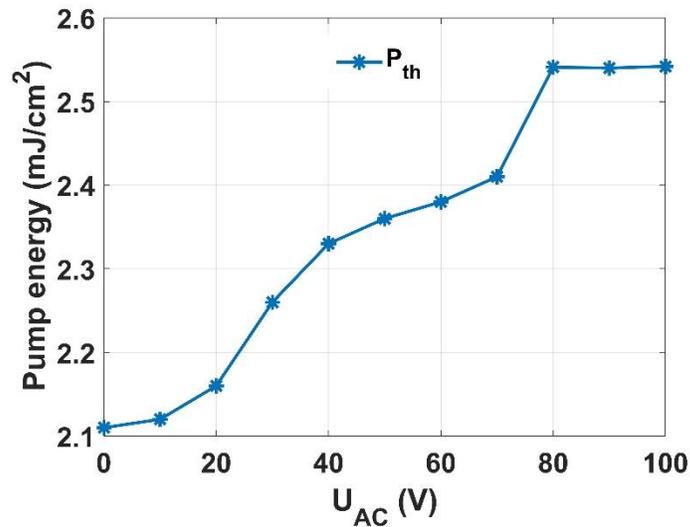

Fig. S5. Threshold of the WGM laser as a function of $U_{AC}$ when $P_{in} = 2.4$ mJ/cm$^2$ and $D = 75$ μm

## 6. Influence of pump power on sample temperature

To investigate the effect of pumping on sample temperature, a thermal infrared camera (FLIR E6-XT, −20°C~550°C) was used to measure the temperature profile of the sample without the heating device every 2 minutes, as shown in Fig. S6. During the measurement, the repetition frequency of the 532 nm pulse laser was set to 5 Hz (the highest value of our pump) and the energy density was set above the threshold of WGM lasing (2.5mJ/cm$^2$), the sample was pumped shot by shot. As shown in Figs. S6(a)-(c), the overall temperature of the sample is still maintained at around 25°C (ambient temperature) after pumping continuously for 1, 4 and 7 minutes, which proves that the thermal effect of the pump has a negligible influence on the experimental results. Fig. S6(d) further shows the variation of the sample temperature with the pumping time in 10 minutes, and the temperature keeps almost unchanged.

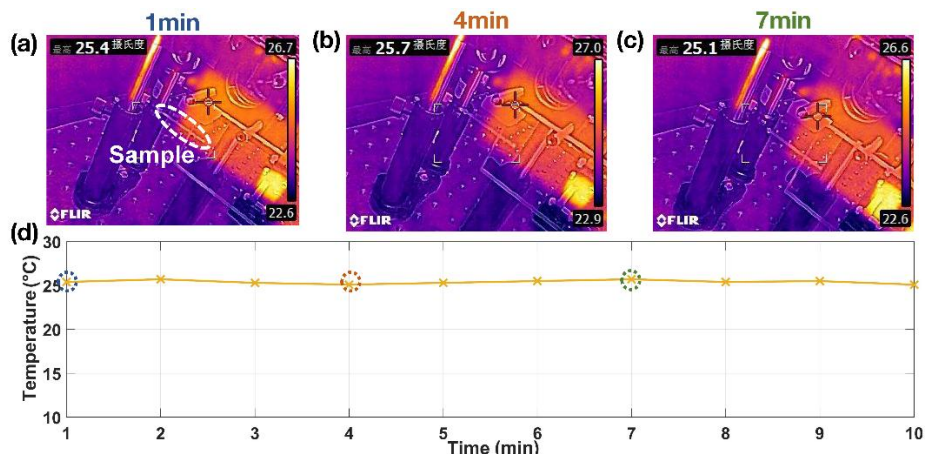

Fig. S6. The thermal images of the sample when pumped by the nanosecond pulses. (a)-(c) The thermal images at different pumping time. (d) The temperature of the sample as a function of pump time